 \documentclass[preprint2]{aastex}

\newcommand{\lesim}{\,\raisebox{-0.4ex}{$\stackrel{<}{\scriptstyle\sim}$}\,}

\shorttitle{HLX-1's accretion disk during outburst}
\shortauthors{Webb et al.}

\begin{document}


\title{Optical variability of the accretion disk around the intermediate\\
   mass black hole ESO 243-49 HLX-1 during the 2012 outburst}


\author{N. A. Webb\altaffilmark{1,2}, O. Godet\altaffilmark{1,2}}
\affil{1 Universit\'e de Toulouse; UPS-OMP; IRAP;  Toulouse, France}
\affil{2 CNRS; IRAP; 9 av. Colonel Roche, BP 44346, F-31028 Toulouse cedex 4, France}\email{natalie.webb@irap.omp.eu}
\author{K. Wiersema\altaffilmark{3}}
\affil{3 University of Leicester, University Road Leicester, LE1 7RH}
\author{J.-P. Lasota\altaffilmark{4,5,6}}
\affil{4 Institut d'Astrophysique de Paris, UMR 7095, CNRS, UPMC Universit\'e Paris 06, 98bis Boulevard Arago, 75014 Paris, France}
\affil{5 Nicolaus Copernicus Astronomical Center, ul. Bartycka 18, PL-00-716 Warszawa, Poland}
\affil{6 Astronomical Observatory, Jagiellonian University, ul. Orla 171, 30-244 Krakow, Poland }
\author{D. Barret\altaffilmark{1,2}}
\author{S. A.  Farrell\altaffilmark{7} }
\affil{7 Sydney Institute for Astronomy, School of Physics, The University of Sydney, NSW 2006, Australia}
\author{T. J. Maccarone\altaffilmark{8}}
\affil{8 Department of Physics, Box 41051, Texas Tech University, Lubbock TX 79409-1051, USA}
\and
\author{M. Servillat\altaffilmark{9}}
\affil{9 Laboratoire AIM (CEA/DSM/IRFU/SAp, CNRS, Universit\'e Paris Diderot), CEA Saclay, Bat. 709, 91191 Gif-sur-Yvette, France}




\begin{abstract}
We present dedicated quasi-simultaneous X-ray ({\em
  Swift}) and optical ({\em Very Large Telescope} (VLT), V- and R-band) observations of the intermediate mass black hole candidate
\objectname[ESO 243-49 HLX-1]{HLX-1} before and during the 2012
outburst.  We show that the V-band magnitudes vary with time, thus proving that a portion of the observed emission originates in the accretion disk. Using the first quiescent optical observations of \objectname[ESO 243-49 HLX-1]{HLX-1}, we show that the stellar population surrounding \objectname[ESO 243-49 HLX-1]{HLX-1} is fainter than V$\sim$25.1 and R$\sim$24.2.  We show that the optical emission may increase before the X-ray emission consistent with the scenario proposed by \cite{laso11} in which the regular outbursts could be related to the passage at periastron of a
star circling the intermediate mass black hole in an eccentric orbit, which triggers mass transfer into a quasi-permanent accretion
disk around the black hole. Further, if there is indeed a delay in the X-ray emission we estimate the mass-transfer delivery radius to be $\sim$10$^{11}$ cm.
\end{abstract}





\section{Introduction}
\label{sec:intro}

Two varieties of black hole (BH) have been extensively observed: stellar mass ($\sim$3-20 M$_{\odot}$) BHs and supermassive
($\sim$ 10$^{6-9}$ M$_{\odot}$) BHs present in the cores of most
galaxies. It is believed that stellar mass BHs are formed from the
collapse of massive stars \citep[e.g.][]{frye03}, but it is not yet
clear how supermassive ones are formed. One model proposes that they
are formed from the mergers of smaller mass ($\sim$10$^{2-5}$ M$_{\odot}$) BHs, the so-called intermediate mass black
holes \citep[IMBH, e.g.][]{mada01}. However, the observational
evidence for these BHs has been, until recently, fairly weak.

One very good candidate is a 2XMM X-ray catalogue source \citep{wats09}: 
\objectname[ESO 243-49 HLX-1]{2XMM J011028.1-460421}, serendipitously
discovered as an off-nuclear X-ray source apparently associated with
the galaxy \objectname[]{ESO 243-49}, 95 Mpc away
\citep{farr09}. Follow-up spectroscopy of the optical counterpart
confirmed the association \citep{wier10}. Using the maximum unabsorbed
X-ray luminosity of 1.1 $\times$ 10$^{42}$ erg s$^{-1}$
\citep[0.2-10.0 keV,][]{farr09} and the conservative assumption that
this value exceeds the Eddington limit by at most a factor of 10,
implies a minimum mass of 500 M$_\odot$. Modelling and Eddington
scaling of the X-ray data indicate a mass of the order
10$^4$ M$_\odot$ \citep{gode12,davi11,serv11} and recent radio
observations provide a mass upper limit of 9 $\times$ 10$^4$
M$_\odot$ \citep{webb12}. This source, now
known as \objectname[ESO 243-49 HLX-1]{ESO 243-49 HLX-1} or HLX-1 for
short, is therefore a very good candidate IMBH.

Regular monitoring of HLX-1 with {\em Swift} has revealed flux changes
by a factor of 50 in conjunction with simultaneous spectral changes in
the same way as Galactic BH X-ray binaries \citep{gode09,serv11}, thus
strengthening the case for an accreting BH in HLX-1. It has become
evident that HLX-1's X-ray variability follows a fairly distinct
pattern over $\sim$1 yr \citep{laso11,gode12,gode12b}. The FRED-type
(Fast Rise Exponential Decay) light curve of \objectname[ESO 243-49
  HLX-1]{HLX-1} outbursts is sometimes observed in Galactic Black Hole
Low Mass X-ray Binaries (BHLMXBs).  However, \citet{laso11} showed
that the \objectname[ESO 243-49 HLX-1]{HLX-1} outbursts cannot be
explained by the same mechanism: the Disk Instability Model
\citep[DIM, see][for a review]{laso01}.  In the framework of the DIM,
the peak outburst luminosity requires the outer disk radius to be
$>10^{13}\rm\ cm$, incompatible with the observed decay time. For such
a radius, the viscous decay time is $>100$\,yr, whereas according to
observations it is just a few months. The (apparently) periodic outbursts
of HLX-1 can therefore not be attributed to a thermal-viscous disk
instability.

As shown by \citet{laso11} an evolved (AGB - Asymptotic Giant Branch)
star on an eccentric orbit around the $\sim10^4\, M_{\odot}$ BH,
periodically delivering part of its mass to a permanently present
accretion disk, can provide a viable explanation of the HLX-1
outbursts. This scenario has to take into account the outburst decay
time which requires a disk radius (or a mass-transfer delivery radius)
of $<10^{12} \rm\ cm$, which implies a very eccentric orbit of the
putative donor star. In this model the impulsive increase in mass
transfer leads to a rise that is a
fraction of the viscous time ($\lesim 0.1 t_{\rm vis}$) and then
decays on $\sim t_{\rm vis}$. The passage of the star at periapse may
also lead to the excitation of waves in the disk that will enhance
angular momentum transport \citep{Spruit-87}.  If such wave propagation
mechanisms are at work during the rise to outburst of HLX-1, the
characteristic propagation velocity might be similar to that of the
heating front according to the DIM ($\sim \alpha c_s$, where $\alpha
\sim 0.1$ and $c_s$ is the speed of sound). In such a case
the outburst always starts in the outer disk and propagates inwards.

In BHLMXBs the X-ray emission has been observed to lag the optical
during outburst \citep{oros97,jain01,buxt04,wren01} but, contrary to
na\"ive expectations, they are not thought to be produced from the
outbursts starting at the outer disk edge and propagating inwards
\citep{Hameury-97}. The characteristic BHLMXB rise timescales ($\sim$days) and the fact that delays between the rise of individual optical
bands \citep[e.g.][]{oros97} are also sometimes observed, suggest that
the X-ray/optical tracks the size of the inner truncation radius of
the quiescent disk. In LMXBs the heating fronts triggered by the
thermal/viscous instability cannot propagate into the inner advection
dominated accretion flow \citep[ADAF, e.g.][]{Dubus-01} which is
invaded by the accretion disk on a viscous time. Typically the radii
of the ADAF region corresponding to delays of a few days are $\sim10^4 R_S$ (where $R_S=2GM/c^2$), i.e. $\sim10^{10}$ cm.  The case of
HLX-1 is different since its outbursts are not due to a
thermal/viscous instability but to an impulsive increase of the mass
transfer in the outer disk. Therefore, in principle, an observed delay
between the rise of X-rays and the optical emission can be attributed
to the distance between the two emitting regions.

In this paper we examine dedicated quasi-simultaneous {\em
  Swift} X-ray and  {\em Very Large telescope} (VLT, V- and R-band) optical observations of
\objectname[ESO 243-49 HLX-1]{HLX-1} before and during its 2012
outburst \citep{gode12} with the aim of constraining
the properties of the rise-to-outburst mechanism. We also compare Gemini
R-band data from the X-ray plateau phase and beginning of the X-ray
decay, taken during the 2011 outburst. We examine the lightcurves for any evidence of a delay between
the rise of X-rays with respect to the optical, which may provide information about the propagation of
the perturbation at the outburst origin, as well as the disk
structure.

\section{Observations}

\subsection{Optical data}

We observed \objectname[2XMM J011028.1$-$460421]{HLX-1} under the ESO
program (089.D-0360(A), PI: Webb) on 7 nights from 2012 August 11 to
2012 September 8, using the {\em FORS2} instrument on the VLT, see
Table~1. We employed two filters, the `v high' filter, with an
effective wavelength of 557 nm and a Full Width at Half Maximum (FWHM)
of 123.5 nm and the `R special' which has an effective wavelength of
655 nm and a FWHM of 165 nm\footnote{see:
  http://www.eso.org/sci/facilities/paranal/
  instruments/fors/doc/VLT-MAN-ESO-13100-1543\_v91.pdf}.  The
observations were carried out in service mode.  Due to technical
problems with the MIT CCDs, the first three observations were carried
out with the E2V CCDs (2$\times$2 binning, 0.25''/pixel). The E2V CCDs
provide much higher response in the blue compared to the MIT CCDs, but
suffer from strong fringing above 650 nm and thus had an effect on the
`R' observations. The latter 4 observations were taken with the MIT
CCDs (2$\times$2 binning, 0.25''/pixel).   The seeing values, as
determined from measuring the FWHM of stars in the images are given in
Table~\ref{tab:obs}.  The sky was always clear, but not necessarily
photometric.  We used {\sc iraf} version 2.16 \citep{tody86,tody93} to
reduce and analyse the data. The raw images were trimmed, bias
subtracted and then flatfielded using the sky flats taken closest in
time. Some residual fringing of the order 3-4\% remained after this
step and was slightly higher for the R-band observations taken with
the E2V CCDs. The observations on any one night in any particular
filter were divided into two equal exposures to avoid saturating the
host galaxy emission in the \objectname[2XMM J011028.1$-$460421]{HLX-1} region. We then added the two
exposures taken in the same filter on each night, to improve the
signal to noise ratio.

\begin{table}[h]
\begin{center}
\caption{Optical observations made with the VLT }
\begin{tabular}{ccccc}
\tableline\tableline Date  & MJD & Filter & Exp. & Seeing\\ (2012) & &
& Time (s) & ('')\\ \tableline 11 Aug. & 56150.38 & v high & 2
$\times$ 390 & 1.0 \\ 11 Aug. & 56150.37 & R special & 2 $\times$ 255
& 0.9\\ 14 Aug. & 56153.20 & v high & 2 $\times$ 390 & 0.8 \\ 14 Aug.
& 56153.20 & R special & 2 $\times$ 255 & 0.8\\ 16 Aug.  & 56155.21 &
v high & 2 $\times$ 390 & 0.9\\ 16 Aug. & 56155.21 & R special & 2
$\times$ 255 & 0.8 \\ 19 Aug. & 56158.25 & v high & 2 $\times$ 390 &
1.3 \\ 19 Aug. & 56158.24& R special & 2 $\times$ 255 & 1.0 \\ 21 Aug.
& 56160.34 & v high & 2 $\times$ 390 & 0.8 \\ 21 Aug.  & 56160.34 & R
special & 2 $\times$ 250 & 0.6 \\ 23 Aug. & 56162.30 & v high & 2
$\times$ 390 & 1.3 \\ 23 Aug.  & 56162.30& R special & 2 $\times$ 250
& 1.3 \\ 8 Sep. & 56178.34 & v high & 2 $\times$ 390 & 0.9 \\ 8 Sep. &
56178.33& R special & 2 $\times$ 250 & 1.0\\ \tableline
\end{tabular}
\label{tab:obs}
\end{center}
\end{table}

\objectname[2XMM J011028.1$-$460421]{HLX-1} is immersed in the diffuse
emission from the galaxy ESO 243-49 in all of the images. This must be
subtracted to determine reliable flux values of \objectname[2XMM
  J011028.1$-$460421]{HLX-1}. We explored six different
methods. Initially we tried Point Spread Function (PSF) matched image
subtraction with spatially varying kernels, using the ISIS2 code
\citep{alar00}.  However, this did not give sufficient control over
the stars selected for the PSF fitting, which caused large
residuals. We then tried the same method, but with the HOTPANTS
codes\footnote{http://www.astro.washington.edu/users/becker/hotpants.html},
which gave more control over the stars selected and over the
photometric errors. However, the proximity of \objectname[2XMM
  J011028.1$-$460421]{HLX-1} to the very brightest parts of the host
galaxy (close to non-linear count values) and the complex changes in
host galaxy brightness on small spatial scales, results in systematic
residuals in the subtracted frames (particularly along the long
symmetry axis of the galaxy and near the nucleus) that strongly
influence the \objectname[2XMM J011028.1$-$460421]{HLX-1} flux
determination. Changing the image section sizes and positions did not
sufficiently negate this problem.  

We also tried elliptical isophote fitting but were unable to fit and
remove the diffuse emission effectively, as strong deviations of
elliptical symmetry were apparent. We then tried two methods explored
by \cite{sori12}, first by using PSF-convolved physical model fits to
the host galaxy as a whole, using the GALFIT codes \citep{peng10} and
subtracting this model. However, we found that a large number of model
components were required to eliminate strong residuals near
\objectname[2XMM J011028.1$-$460421]{HLX-1}, which in turn lead to
large systematic errors associated with model degeneracy. 

We next attempted surface-fits to the local brightness profile and
subtracted these from the diffuse emission, as explored in
\cite{sori12}.  We excised postage stamp images containing
\objectname[2XMM J011028.1$-$460421]{HLX-1}, with varying centroid
coordinates and stamp sizes. We fitted polynomial surfaces to the
light profile, excluding a circular region around \objectname[2XMM
  J011028.1$-$460421]{HLX-1}. We used several radii of exclusion as
well as a range of polynomial order values. However, both background
residuals and source flux varied strongly with small variations in
fitted image stamp and exclusion region.  This is likely to be due to
the high order polynomial and small stamp size to adequately trace the
strong variations in galaxy brightness (steep flux gradient) on small
spatial scales.  We therefore attempted to remove the steepest flux
gradients by exploiting local symmetry, subtracting an image of the
galaxy mirrored in the long axis, before performing local surface
brightness fits in postage stamps ($\sim$15\arcsec
$\times$15\arcsec\ in size) around \objectname[2XMM
  J011028.1$-$460421]{HLX-1}. We then required much lower polynomial
orders for the background fit, and residuals were robust to small
changes in fit regions. This revealed \objectname[2XMM
  J011028.1$-$460421]{HLX-1} clearly except for observations on August
11 and 14.  For these observations, we added the data from the two
nights in each filter.  \objectname[2XMM J011028.1$-$460421]{HLX-1}
could then be identified.

We also reduced and analysed the {\em Gemini} r' filter (562-698 nm)
presented in \cite{farr13} and taken on four nights between 2011
August 31 and 2011 October 6.  The detector pixel scale is 0.146'' per
pixel and we used a 9-point dither pattern repeated twice.  We reduced
and analysed these data in the same way as the {\em VLT} observations.

As the {\em VLT} observations were not always carried out in
photometric conditions, we give relative fluxes (compared to a
non-variable field star) in Fig.~\ref{fig:lc}. These fluxes were
calculated using the {\em phot} task in {\sc iraf} by means of
aperture photometry. We used an aperture of 1.5 FWHM to extract the
source counts and employed the {\sc iraf} task {\em mkapfile} to
correct for aperture losses. We also used a larger annulus around the
extraction region and within the region of low and stable background
to determine a background level. We utilized the instrumental
zeropoint values calculated for the night our observations were made,
and given on the FORS2
webpages\footnote{http://archive.eso.org/bin/qc1\_cgi?action=qc1\_browse\_
  table\&table=fors2\_photometry}, along with the extinction and the
colour corrections.  To verify the reliability of the fluxes and
magnitudes extracted, we checked that the magnitudes of 9 field stars
did not vary over all the observations to within the magnitude
errors. However, it should be noted that the R-band errors are
slightly larger than the V-band errors, due in large part to the first
three observations taken with the blue CCD that shows strong fringing
in this band. We included stars as faint as \objectname[2XMM
  J011028.1$-$460421]{HLX-1} and which were immersed in the diffuse
galaxy emission. These objects were reduced in the same manner as
\objectname[2XMM J011028.1$-$460421]{HLX-1}, and include one star that
falls less than 4\arcsec\ from \objectname[2XMM
  J011028.1$-$460421]{HLX-1}, see Fig.~\ref{fig:image}, which was
therefore extracted from the same small postage stamp region. Only one
object in the field was found to vary between observations, and this
was \objectname[2XMM J011028.1$-$460421]{HLX-1}.  The variability of
\objectname[2XMM J011028.1$-$460421]{HLX-1} can be seen in
Fig.~\ref{fig:lc}, along with the relative R-band flux of a field
star, to show the reliability of the data reduction and analysis.

\subsection{X-ray data}

The {\em Swift X-ray Telescope} (XRT) Photon Counting data were
processed using the {\em Swift}-XRT light-curve generator web
interface, binned by observation \citep{evan09}. No X-ray source was
detected at the position of \objectname[2XMM
  J011028.1$-$460421]{HLX-1} for the observations made before MJD
56160. For these observations we calculated the 3$\sigma$ upper limit,
as explained in \cite{evan09}.

\section{Results and discussion}

\begin{figure*}
\includegraphics[angle=0,scale=.39]{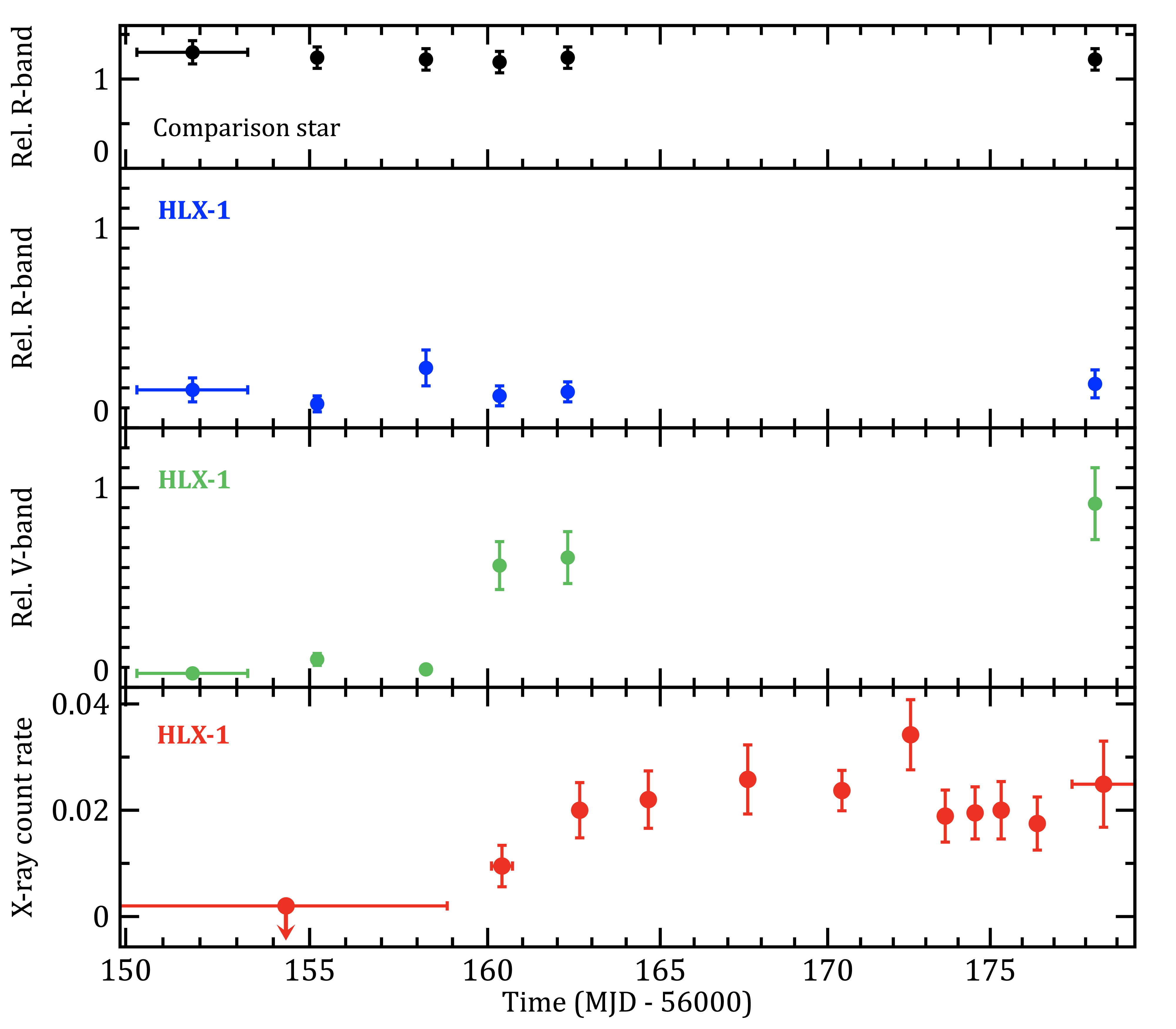}
\caption{The uppermost panel shows the relative `R special' flux of a
  field star and the comparison star used for the HLX-1 lightcurves.
  The middle two panels show the relative R- and V-band flux between
  HLX-1 and a comparison star. The errors are the percentage errors
  calculated from the photometry. The lower panel shows the {\em
    Swift} X-ray counts (0.2-10.0 keV) measured for HLX-1 over the
  same period. X-ray error bars are 1$\sigma$. The first three X-ray
  points are an average upper limit for the three observations. Wide
  time errors indicate that data on more than one day has been added.}
\label{fig:lc}
\end{figure*}

The X-ray and optical lightcurves can be seen in Fig.~\ref{fig:lc}.
No X-ray source is detected before the observation on MJD 56160, which
is consistent with the count rate observed in the low hard state
\citep[e.g.][]{gode12}. The X-ray count rate appears to commence a
rise on MJD 56160, however this point (0.0095$\pm$0.0039 ct s$^{-1}$),
which is shown in Fig.~\ref{fig:lc} with its error bar at the
1$\sigma$ level, has a count rate consistent with the low hard state
count rate if we consider an error bar at the 2$\sigma$ level and with
zero counts if we consider the 3$\sigma$ error bar
\citep{gode12b}. Therefore, there is only a hint that the outburst
began on this day and it is only two days later (MJD 56162) that we
can confidently say that the outburst began in X-rays, as discussed in
\cite{gode12} and \cite{kong12}.  Fitting the X-ray data-points, the
outburst maximum (peak) is reached around MJD 56164. The observations
until MJD$\sim$56178 are consistent with a constant count rate (within
the errors, $\chi^{\scriptscriptstyle 2}_{\scriptscriptstyle \nu}$ =
0.73, for 9 degrees of freedom). We therefore observe a plateau phase
of at least two weeks, as observed in previous outbursts
\citep{gode12}.

The V-band flux also rises on MJD 56160. However, contrary to the
X-rays, the relative V-band flux observed on that day (0.61$\pm$0.24,
3$\sigma$ error) is not consistent with the quiescent value
(0.09$\pm$0.04, 3$\sigma$ error) at  the 3 $\sigma$ level.  Thus the
optical rise may be interpreted to have commenced on MJD 56160,
potentially before the X-ray, which would be indicative of an
outside-in outburst and supporting the scenario proposed by
\cite{laso11}, see Sec.~\ref{sec:intro}.  

The R-band does not show a rise on the same timescale.  This may be
because the host stellar population \citep{farr12} dominates in the
R-band, whereas the disk dominates the V-band, or that there is some
other cool emission, for instance from a nebula \citep{sori13}. This
is supported by the V-band magnitude which is very faint in the low
state and increases with the outburst, see
Table~\ref{tab:outburst_mags}, whereas the R-band is brighter
initially.  

\begin{figure}
\includegraphics[angle=0,scale=.3]{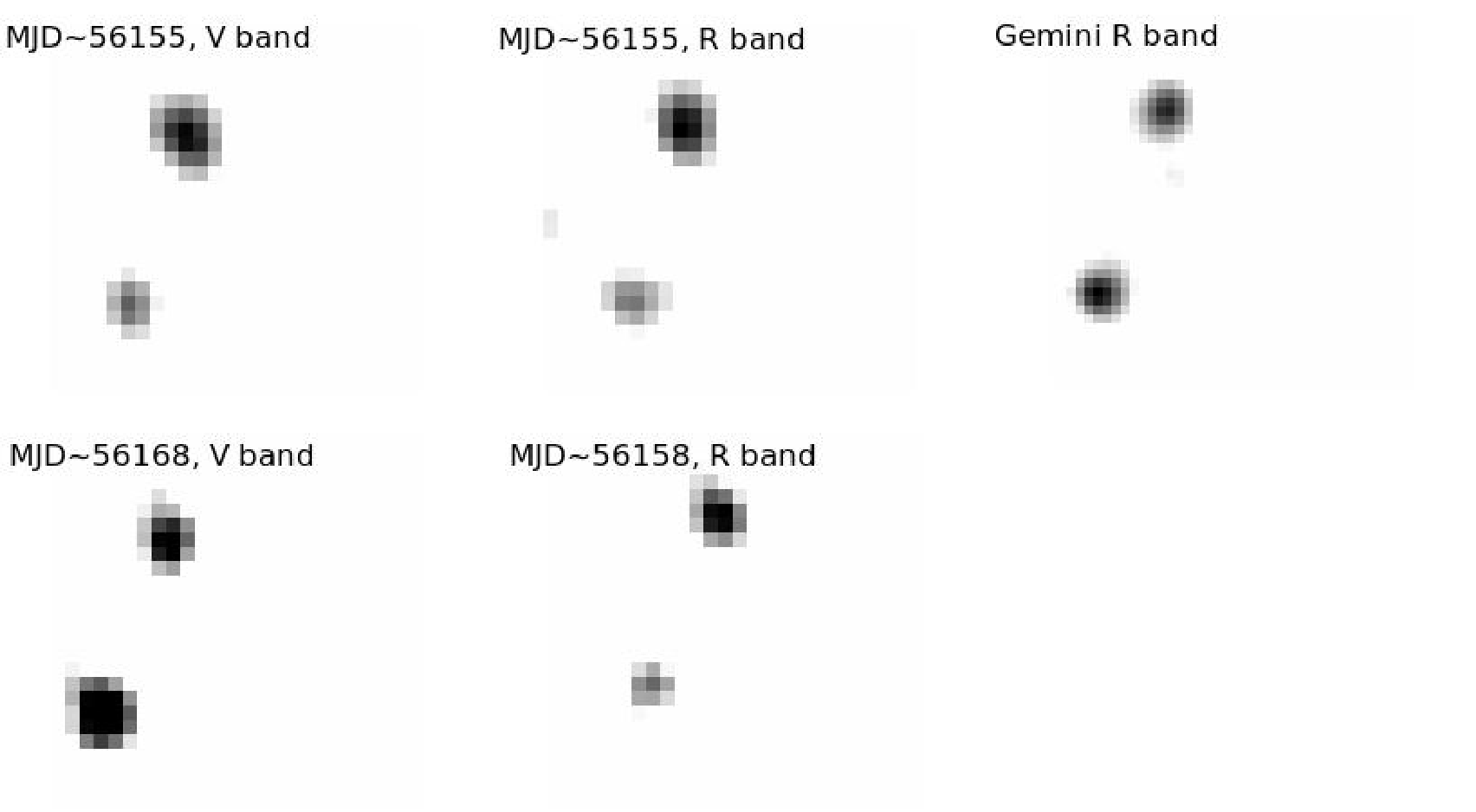}
\caption{Five images, each showing a comparison star (upper object)
  and HLX-1 (lower object). The left panels show the {\em VLT} V-band
  observations. Upper left: before the 2012 outburst commences. Lower
  left: $\sim$15 days after the 2012 outburst peaks. The middle panels
  show the {\em VLT} R-band observations. Upper middle: before the
  2012 outburst commences.   Lower middle: $\sim$5 days before the
  2012 outburst peaks. Upper right: {\em Gemini} r' band image,
  $\sim$15 days after the 2011 outburst peak.  All images are scaled
  to show the comparison star with a similar count rate.}
\label{fig:image}
\end{figure}

The flux remains roughly constant in the {\em VLT} R-band
observations.  \cite{farr13} also see little evidence for variability
in the R-band, using {\em Gemini} observations taken slightly later in
the 2011 outburst (starting two weeks after the peak of the X-ray
outburst that year).  Our analysis of the same data also reveals no
evidence for variability in relative flux compared to the same
non-variable field star used for the {\em VLT} data, to within the
observational errors. \objectname[2XMM J011028.1$-$460421]{HLX-1}'s
relative flux in the {\em Gemini} data appears approximately 1
magnitude brighter in the R-band, compared to the low level detected
in the majority of the {\em VLT} R-band observations, see
Fig.~\ref{fig:image} and Table~\ref{tab:outburst_mags}. However, at
the 3 $\sigma$ level, these two values are compatible.  We verified
that the observed difference in magnitude was not due to variability
in the comparison star, by comparing it to a second non-variable field
star. No variability is detected in the original comparison star in
any of the {\em VLT} or {\em Gemini} observations.  

We have estimated the R- and the V-band magnitudes for the {\em VLT}
and {\em Gemini} data and they are presented in
Table~\ref{tab:outburst_mags}. The variability can also be visualised
in Fig.~\ref{fig:image}. We compare our estimated (because the
conditions were not always photometric) magnitudes with previous
observations, taken at different times through the 2009-2012
outbursts.  It should be noted, however, that the optical observations
made before MJD 56160 are the first to be made during the low-hard
state. The very low magnitude measured indicates that any host stellar
population should have a maximum V-band magnitude of $\sim$25.1
(taking into account the V-band errors).  Due to the
limited number of filters used in this study, it is difficult to
constrain the nature of the underlying stellar population and the
contribution of the accretion disc, as previously done by
\cite{farr12}.  However, the bright state data are consistent with the
irradiated disc and an underlying stellar population presented in
\cite{farr12}.

\begin{table}
\begin{center}
\caption{Magnitude estimates compared at similar times over different
  outbursts}
\begin{tabular}{p{3.2cm}p{1.0cm}p{0.8cm}p{1.5cm}}
\tableline\tableline Magnitude & Days & Year & Telescope\\ & after
peak &  & \\ \tableline R$_{AB}$=24.5$\pm$0.3$^*$ & $\sim$365 & 2012 &
VLT \\ R$_{vega}$=24.71$\pm$0.40$^\clubsuit$ & $\sim$90 & 2010 & VLT
\\ \tableline R$_{AB}$=23.5$\pm$0.3$^*$ & $\sim$15 & 2011 & Gemini
\\ R$_{cousin}$=23.80$\pm$0.25$^\triangle$ & $\sim$10 & 2009 &
Magellan \\ \tableline \tableline V$_{AB}$=25.4$\pm$0.3$^*$ &
$\sim$365 & 2012 & VLT \\ V$_{vega}$=24.79$\pm$0.34$^\clubsuit$ &
$\sim$80 & 2010 & VLT \\ \tableline V$_{AB}$=24.3$\pm$0.2$^*$ &
$\sim$-2 & 2012 & VLT \\ V$_{cousin}$=24.5$\pm$0.3$^\triangle$ &
$\sim$10 & 2009 & Magellan \\ \tableline V$_{AB}$=23.6$\pm$0.2$^*$ &
$\sim$15 & 2012 & VLT \\ V$_{AB}$=23.83$\pm$0.08$^\diamondsuit$ &
$\sim$25 & 2010 & HST \\ \tableline
\end{tabular}
\label{tab:outburst_mags}
\tablecomments{$^*$This work\\ $^\triangle$\cite{sori10}
  \\ $^\clubsuit$\cite{sori12}\\ $^\diamondsuit$\cite{farr12}
  \\ V$_{AB}$-V$_{Vega}$=0.02, R$_{AB}$-R$_{Vega}$=0.21 \citep{blan07}
  \\ V$_{cousin}$=V$_{AB}$+0.044, R$_{cousin}$=R$_{AB}$-0.055
  \citep{frei94}}
\end{center}
\end{table}

As shown above, there is a possibility that the V-band rises before
the X-ray. Due to the sampling timescale of $\sim$2 days, we can give
an upper limit on the possible delay in the X-rays of $<$2 days.  We
estimate the mass transfer delivery radius that a delay of the order
of 1 day corresponds to in the case of HLX-1.

If one assumes that the impulsive increase in the rate at which matter
is delivered to the disk propagates as a density contrast, the viscous
propagation time can be written as
\begin{equation}
t_{\rm prop}\approx \frac{R(\Delta R)}{\nu}= \delta t_{\rm  vis},
\end{equation}
where $\Delta R$ is the width of the density contrast, $\delta=\Delta
R/R$, $\alpha < 1.0$, $\nu=\alpha c_s^2 /\Omega$ is the kinematic
viscosity coefficient and $\Omega_K=\sqrt{{GM}/{R^3}}$ is the
Keplerian angular velocity \citep{Hameury-97}. Hence the distance the
density contrast will go through is
\begin{equation}
R_{\rm tr}\approx 5.8 \times 10^{9}\, t_d^2\,
\delta_{-2}^{-1}\,\alpha^2_{0.1}\,T^2_7 M_4^{-1}\,\rm cm,
\end{equation}
where $t_d$ is the delay time in days and $\delta_{-2} =
\delta/10^{-2}$, $\alpha_{0.1}=\alpha/0.1$, $T=10^{7}T_7$ K is the
midplane temperature and $M=(10^4 M_{\odot}) M_4$ the BH mass. Such a
distance ($\sim 2R_S$) is too short to constrain the properties of the
accretion flow in HLX-1 if the rise to outburst is driven by
viscosity. One could increase $R_{\rm tr}$ by making the density
contrast sharper but since the assumed temperature ($10^7$\,K) is
already too high for an optical emitting region such playing with
numbers is of little interest.

However, as mentioned above, the required HLX-1 disk size
makes the viscous character of the rise to outburst extremely unlikely
and the wave-propagation mechanism is certainly to be preferred. The
sound crossing distance corresponding to 1 day is
\begin{equation}
R_{\rm sound}\approx 3\times 10^{11}\, T^{1/2}_5\, t_d \rm \,cm,
\end{equation}
where $T=10^{5}T_5$ K. This radius of $\sim3\times 10^{11}$ cm can
then be taken as an upper limit and might correspond to the distance
between the optical and X-ray emitting regions if the density contrast
propagates through waves. This radius is 2-3 magnitudes smaller than
the outer disk radius determined in \cite{farr13}.

One can conclude, therefore, that if the putative observed delay
between the onset of \objectname[2XMM J011028.1$-$460421]{HLX-1}'s
optical and X-ray outbursts results from the propagation of the
density contrast created by an impulsive increase of mass-transfer
into an accretion disk, this propagation cannot be viscous but must be
mediated by waves in agreement with the suggestion of \citet{laso11}.

\acknowledgments

We thank the anonymous referee, whose comments greatly improved
the paper.  JPL was supported by grants from the CNES and the Polish
National Science Center (UMO-2011/01/B/ST9/05439).  SAF receives an
Australian Research Council Postdoctoral Fellowship (DP110102889). MS acknowledges support from the CNES. We thank
N. Gehrels for granting us the {\em Swift} observations. This work
used data supplied by the UK {\em Swift Science Data Centre} at
the University of Leicester.  Based on observations made with ESO
Telescopes at the Paranal Observatory under programme ID 089.D-0360(A)
and on observations obtained at the Gemini Observatory.


\clearpage



\clearpage









\clearpage

\end{document}